\documentclass[11pt]{article}
\usepackage{epsfig,graphics}
\begin{document}

\null
\bigskip
\bigskip
\bigskip
\centerline{\bf \LARGE Description of the Damn Yankee Controller (DYC)}
\bigskip

\centerline{\Large Steve Bracker}
\medskip
\centerline{\Large Box 1290, Enderby, British Columbia V0E\,1V0 \ Canada}
\medskip
\centerline{\Large s\_bracker@hotmail.com}
\medskip
\bigskip
\centerline{\Large Sten Hansen}
\medskip
\centerline{\Large Fermilab, Box 500, Batavia, Illinois 60510 \ USA}
\medskip
\centerline{\Large hansen@fnal.gov}

\bigskip
\bigskip
\bigskip
\centerline{\bf ABSTRACT}
\bigskip

\centerline{\parbox{5.8in}{
Versions of the Damn Yankee Controller (DYC) have been used to read out
digitizers on the Fermilab E--791, E--835, FOCUS, SELEX, and KTeV experiments.
The DYC accepts 16-bit data and control signals from 10 MHz 
Emitter Coupled Logic (ECL) PORT digitizing
modules such as LeCroy CAMAC PCOS latches and FERA Analog to Digital 
Converters (ADCs). Data is packed into
First In First Out (FIFO) memories as 32-bit longwords. Complete events
including data, a leading
word count, and an Event Synchronization Number are transmitted to a data
destination. The DYC described here was meant to be simple and fast
and was designed and built in a period of three weeks.}}

\bigskip
\bigskip
\bigskip
\leftline{\bf INTRODUCTION}

In the Fermilab E--791 charm hadroproduction experiment, 
there are three Damn Yankee Controllers (DYCs) in the digitizing logic. 
Each is associated with a particular digitizing system:
\begin{enumerate}
\item The LeCroy CAMAC 4300B FERA1 ADC system
\item The LeCroy CAMAC 4300B FERA2 ADC system
\item The LeCroy CAMAC 2731A PCOS LATCH system
\end{enumerate}

The LeCroy PCOS latches digitized 10 planes of proportional wire chambers
in 4 $\mu$S including readout. The LeCroy [1] Fast Encoding and Readout
ADCs (FERAs) digitized two Cherenkov
counters [2], an electromagnetic calorimeter [3], and a hadronic calorimeter
[4] in 30 $\mu$S including readout. The simplicity of CAMAC was retained,
while exploiting the vastly increased speeds of the PCOS and FERA modules
front 10 MHz ECLPORT with a simple token passing, sequential readout. The
increased speed was a key to the recording of 20 billion events with the
E--791 data acquisition system [5] during Fermilab's 1991-1992 fixed target
run. The resulting 50 Terabyte data set was reconstructed using large
computing farms at four sites [6] and yielded 200\,000 fully reconstructed
charmed hadrons. Parts of each of these 20 billion events passed through DYCs
designed and built in May 1991.

Each of the 
three E--791 DYCs stands between a DIGITIZING SYSTEM and a DATABUS leading
to an EVENT FIFO BUFFER (for details of the EFB see references [5] and [7]
and Figure 1). 
THE EFBs were originally used as a Video Data Acquisition System (VDAS).
Data produced by the digitizing system is
received by the DYC and retransmitted over the databus to the Event Fifo
Buffer. More than one data source may deliver data to the EFB over the
databus; thus the data sources must coordinate their use of the databus.

In E--791, one databus (to EFB7) is shared by a DYC serving a FERA ADC
system and another DYC serving the PCOS system. A second databus (to EFB8)
is shared by a DYC serving a FERA ADC system and a camac Smart Crate
Controller (SCC) [8].

\begin{figure}[t!]
\centerline{\epsfxsize 3.375 truein \epsfbox{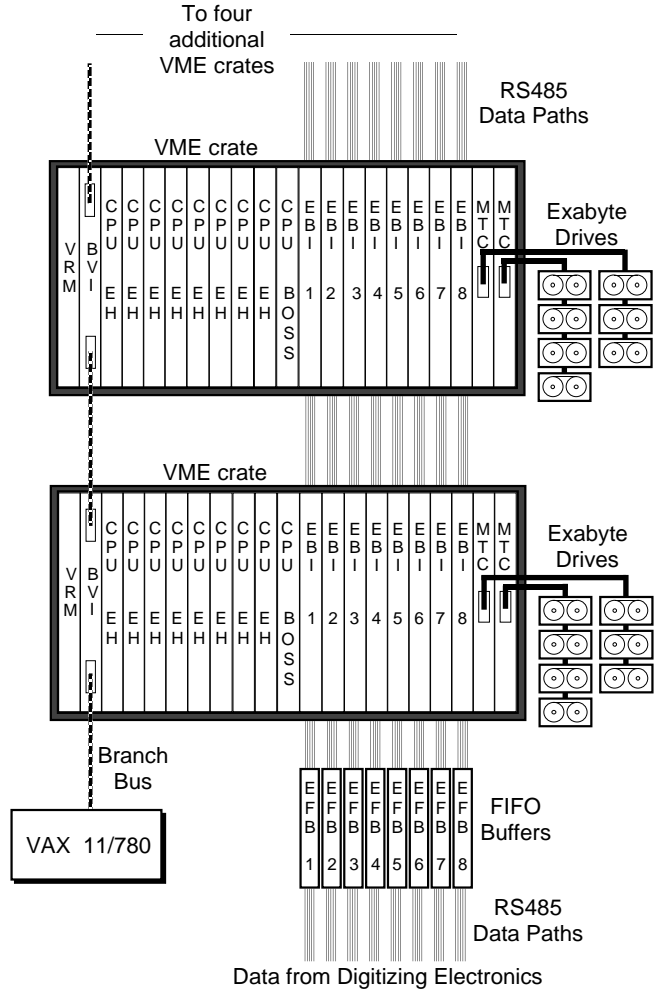}}
\caption[]{A schematic of the VME part of the E--791 DA system [5]. Two complete
VME crates are shown, with the Event Fifo Buffers (EFBs) and data paths from
the digitizers at the base. Each VME crate was attached to each FIFO to form a
$6\times 8$ switching matrix. Six events could thus be built in parallel. Each
of the eight FIFOs was attached to two controllers which shared an RS--485 data
path. Of the 16 controllers, three were DYCs.}
\label{schematic}
\end{figure}

\begin{table}[t!]
\begin{center}
\def \longw{longwords}
\caption[]
{E791 Front End Digitization Systems and Read Out Controllers [5].}
\vspace*{3pt}
\renewcommand{\arraystretch}{1.2}
{\footnotesize
\begin{tabular}{llllll} \hline
 System             & Drift        &  \v Cerenkov,    & Silicon Micro-
& Proportional   & CAMAC        \\
                   & Chamber      &  Calorimeter     & vertex Detector
& Wire Chamber   &              \\ \hline
 Digitizer          & Phillips  & LeCroy 4300B     & Ohio State,
& LeCroy 2731A   & LeCroy       \\
                   & 10C6 TDC     & FERA ADC         & Nanometric N339P,
& PCOS Latch       & 4448 Latch,  \\
                   &              &                  & Nanometric S710/810
&                & 4508 PLU,    \\
                   &              &                  & Latches
&                & 2551 Scaler  \\
 Mean Dead Time     & 30 $\mu$s    & 30 $\mu$s        & 50 $\mu$s
      & 4 $\mu$s    & 30 $\mu$s \\
 Pre-Controllers    & none         & 2 LeCroy 4301s   & 81 Princeton Scanners
& 2 LeCroy 2738s & none         \\
 Controller         & FSCC        & Damn Yankee      & Princeton 
& Damn Yankee    & SCC [8]      \\
 No.~of Controllers & 10           & 2                & 2
& 1              & 1            \\
 Channels / System  & 6304         & 554              & 15896
& 1088           & 80           \\
 Event Size to EFB  & 480 \longw   & 160 \longw       & 110 \longw
& 20 \longw      & 11 \longw    \\
 Event Size to Tape & 300 \longw   & 160 \longw       & 110 \longw
& 20 \longw      & 12 \longw    \\
 On Tape Fraction   & 50\%         & 27\%             & 18\%
& 3\%            & 2\%          \\
\hline \end{tabular}}
\end{center}
\end{table}

The Damn Yankee Controller is intended to fulfill the same function
as the previous University of Mississippi Controller (UMC). It...  
\vspace*{-2pt}
\begin{list}{---}{
\setlength{\itemsep}{0pt}
\setlength{\topsep}{0pt}
}
\item{accepts 16-bit data and control signals from ECLPORT digitizing modules 
   (ADCs and PCOS latches)}
\item{latches 4-bit Event Synchronization Number (ESN) for the event}
\item{accumulates the data word count for the event}
\item{reformats the data received into 32-bit longwords}
\item{transmits a header longword (word count and ESN) to a data destination
   such as an Event Fifo Buffer}
\item{transmits the data previously received and reformatted to the data
   destination}
\end{list} 

The DYC can share the data destination (and the bus to it) with other
data transport modules which adhere to the same protocol. In particular, 
it is compatible with Camac Smart Crate Controllers. 

In addition to its initial use on E--791 [9], versions of the DYC have
subsequently been built and used to make discoveries on the Fermilab E--835
[10], FOCUS [11], SELEX [12], and KTeV [13] experiments.

\bigskip
\bigskip
\bigskip
\leftline{\bf DIFFERENCES BETWEEN THE UMC AND THE DYC}

The differences between the DYC and the UMC are:
\begin{enumerate}
\item The DYC provides true token-passing bus arbitration using the same
protocol as the camac Smart Crate Controller.

\item{A variety of RESET mechanisms are provided, including:}
\vspace*{-2pt}
\begin{list}{---}{
\setlength{\itemsep}{0pt}
\setlength{\topsep}{0pt}
}
\item{a dataway Z on the camac crate containing the DYC}
\item{an F9 camac command directed to the DYC}
\item{a reset signal sent to the DYC through a front panel LEMO connector}
\item{pushing a RESET pushbutton switch on the DYC front panel}
\end{list}
A jumper is provided which instructs the DYC to {set, clear} the arbitration
token upon reset.

\item The DYC provides a separate LATCH ESN strobe input to ensure that the ESN 
is always latched when it is valid. Use of this signal or the BUSY IN signal 
to latch the ESN is Programmable Array Logic (PAL) selectable.

\item Five front panel lights indicate various activities and states in the
DYC -- reading data, writing data, fifo has data, output bus enabled,
arbitration token present. 

\item The DYC can transport data to the EFB at 100 ns per 32-bit word, twice
the speed of the UMC. At high data rates, this will lead to increased
total throughput (decreased effective deadtime), though not by anywhere 
near a factor of 2.

\item Improved termination and signal conditioning for the INPUT DATA PORT and 
the INPUT DATA STROBE are provided. This may enhance data integrity, 
especially when the incoming data is transported a long distance.

\item Improved fast-slew drivers are used for the RS-485 outputs (OUTPUT DATA 
PORT and OUTPUT DATA STROBE). The same type ICs are used for data bits and
the output strobe to minimize differential propagation delay. 

\item All of the control logic for the device is subsumed into three PALs, one 
for the camac functions (new), one for the word count, and one for the 
overall sequencing. All sequencing is synchronous with a 25 ns clock; no 
state-machine timings are determined by one-shots.
\end{enumerate}

\bigskip
\bigskip
\bigskip
\leftline{\bf PHASES OF DYC OPERATION}

There are three basic phases to DYC operations:

IDLE \hfill \break
The state logic is held in its "reset" state. No input data is accepted.
No output data is transmitted. No data is present in the DYC fifo memory.

DATA INPUT \hfill \break
Incoming data is accepted by the DYC. The data is stored in the DYC fifo
memory. The incoming word count is accumulated in the DYC. The ESN is
latched in the DYC. No output data is transmitted. 

DATA OUTPUT \hfill \break
Incoming data is not accepted by the DYC. The header longword - word count
and ESN - is output, followed by the data stored in the DYC fifo memory.

The DYC is held in the IDLE state by default. The DYC enters the DATA 
INPUT phase when the asserted edge of BUSY IN is received. The DYC enters 
the DATA OUT phase when the negated edge of BUSY IN is received, though 
no data is actually transmitted until the PERMIT token is received,
authorizing the DYC to drive the output bus. Upon completion of data 
transmission, the DYC releases the output bus and places itself back in 
the IDLE state to await the next event.

No overlapping of phases is permitted. The DYC cannot accept data from the
next event until it is completely finished transmitting the data from the
previous event. A digitizing system attached to a DYC may accept and
digitize another event while the DYC is still transmitting the last one,
but the digitizer cannot transfer any data to the DYC until the previous event
has been completely transmitted by the DYC. The INHIBIT OUT signal is
provided by the DYC to notify digitizers as to when a BUSY IN assertion and
incoming data can be accepted by the DYC. 

\bigskip
\bigskip
\bigskip
\leftline{\bf INPUTS TO THE DYC}

The INPUT DATA PORT (J1) is a 34-pin ribbon cable which accepts data in 
differential ECL (FERABUS) format. The incoming signal pairs are terminated
by 100 ohm resistors across the pair and 820 ohm bias resistors that set the 
input port to 0 when no cable is connected. Four 10125 ICs convert the 
incoming signal pairs to Transistor--Transistor Logic (TTL). 
The 16 TTL signals are connected to both 
halves of the 32-bit wide DATA FIFO BUFFER.

The INPUT DATA STROBE (DSTRB) is a NIM input. It is terminated by 51 ohms to 
ground. An NE521 is used to convert the signal to TTL; the signal can 
optionally be integrated by the NE521 to remove high frequency noise. The 
threshold of the signal can be set by resistors; by default it is set to 0.3 
volts. The TTL data strobe is sent to both halves of the data fifo buffer.

The ESN INPUT PORT (J2) is a 10-pin ribbon cable which accepts a
four-bit Event Synchronization Number in RS-485 format. The incoming 
signal pairs are terminated by 120 ohm resistors across the pair and 1000 
ohm bias resistors that set the inputs to 0 when no cable is connected.
A 96173 is used to convert the ESN to TTL. The four TTL signals are 
connected to a latch (74HC574) which remembers them until they are needed 
when the header longword is output.

The ESN DATA STROBE (LATCH ESN) is a NIM input through a front-panel LEMO
connector. It is terminated by 51 ohms to ground. An NE521 is used to
convert the signal to TTL (CLKESN). It clocks the ESN latch. A short (20 ns
or so) NIM pulse should be transmitted to ESN DATA STROBE to latch the ESN;
it should occur: 
\vspace*{-2pt}
\begin{list}{---}{
\setlength{\itemsep}{0pt}
\setlength{\topsep}{0pt}
}
\item after the trigger for the current event
\item before the BUSY from the current event has been removed  
\item after the header longword for the previous event has been transmitted 
\end{list}
Alternatively, the ESN can be latched on the asserted edge of BUSY IN; the
present DYC PALs do this, so ESN DATA STROBE is unused.

BUSY IN is a NIM input through a front panel LEMO connector. It is
terminated by 51 ohms to ground. It is converted to TTL (BSYIN) by an
NE521. The transition voltage of the input is -0.3 volts. BUSY IN should
normally be negated (0 volts). It should be asserted at least 100 ns prior
to arrival of data from each event. It should be held asserted until all
data from the event has arrived, and should then be negated promptly. The
assertion of BUSY IN permits incoming data to be stored in the data fifo
buffer. The negation of BUSY IN initiates the data output phase of the
DYC.

RESET is a NIM input through a front panel LEMO connector. It is terminated 
by 51 ohms to ground. It is converted to TTL (NIMRST) and used to initiate a 
full reset of the DYC at startup time. When the NIMRST (or any other reset) 
is asserted, the data fifo buffer is cleared, the word count register is 
zeroed, and the DYC's state logic is initialized. The arbitration token is 
either cleared (module not first to transmit) or set (module is first to 
transmit) according to whether the FIRST jumper is installed or not.

PERMIT IN is a TTL input through a front panel LEMO connector. It is pulled
up by 1000 ohms to +5 volts. The signal should normally be held high. A
downgoing pulse of at least 50 ns duration should be transmitted to PERMIT
IN to transfer the arbitration token to the DYC. The DYC will maintain the
token until it has transmitted an event; it then clears its token and passes
a signal to the next module to receive it through PERMIT OUT.

Various CAMAC DATAWAY signals are received from the card edge connector
including the five FUNCTION (F) lines, the MODULE SELECTED (N) line, the
CRATE INITIALIZE (Z) line, and the SECOND STROBE (S2) line. They are
received and decoded by a PLS153, and combined with other reset mechanisms
(NIMRST, BDRST) to produce a SYSTEM RESET (SYSRST) signal to the CY7C330
sequencer.

+6 volt and -6 volt DC power is taken from the camac dataway. Three power 
supplies are generated from these power sources: +5 volts for most of the 
digital logic, -5.2 volts for ECL power and NIM bias voltages, and -2 volts 
for differential ECL terminations. Fuses, transient absorbers, and filter 
capacitors are provided.

\bigskip
\bigskip
\bigskip
\leftline{\bf OUTPUTS FROM THE DYC}

The OUTPUT DATA PORT (J4) is a 64-pin ribbon cable which transmits data to 
the Event Fifo Buffer. The signals are RS-485 signal pairs, driven by eight 
96172 high speed RS-485 drivers. Because the DYC may share the output bus 
with other devices, the drivers are tri-state enabled only when the DYC 
possesses the arbitration token. The output port is not terminated; the two 
ends of the RS-485 bus must be properly terminated externally to the DYC.
Data is transmitted through the OUTPUT DATA PORT at the rate of one word 
every 100 ns. The data is valid for 70 ns of the 100 ns cycle.

The OUTPUT DATA STROBE is a single signal pair transmitted on pins 1 and 2 
of a 10-pin ribbon cable. The strobe is normally high. When data is present 
at the output data port, the strobe is set low for 50 ns and then returned 
high for at least 50 ns. The receiving module should latch the data at the 
RISING (TRAILING) EDGE of the clock. Data at the output data port is 
guaranteed valid for 35 ns before and 25 ns after the rising edge of the 
strobe.

INHIBIT OUT is a TTL signal transmitted through a front panel LEMO 
connector. Normally low, the signal is raised soon after the assertion of 
BUSY IN initiates processing of an event, and is held high until all the data 
from the event has been transmitted and the token has been passed. When the 
DYC is ready to receive another BUSY IN, it lowers the INHIBIT OUT signal 
again.

PERMIT OUT is a TTL signal transmitted through a front panel LEMO 
connector. Normally high, the signal is pulsed down for 50 ns to indicate 
that the DYC has tri-state disabled its output drivers and is passing 
control of the output bus to the module receiving PERMIT OUT.

\bigskip
\bigskip
\bigskip
\leftline{\bf TIMING REQUIREMENTS OF THE DYC}

The digitizer must not assert a new BUSY IN while INHIBIT OUT is asserted by
the DYC. The existing BUSY IN must remain asserted until all data words from
this event have been transferred from the digitizer to the DYC. 

The first active edge of INPUT DATA STROBE from the digitizer must follow
the asserted edge of BUSY IN from the digitizer by at least 25 nanoseconds. 

The negation of BUSY IN by the digitizer must not precede the asserted edge
of the last INPUT DATA STROBE from the digitizer.

The active edges of successive INPUT DATA STROBES must not be closer than
100 nanoseconds to each other.

Another device sharing the output bus with a DYC must not assert PERMIT IN
until it has tri-state disabled all of its drivers on the OUTPUT DATA BUS
and the OUTPUT DATA STROBE. Once having disabled its drivers and passed 
the permit flag to the DYC, it must not reassert its drivers nor emit 
another PERMIT IN to the DYC until the DYC has transmitted a PERMIT OUT.

\bigskip
\bigskip
\bigskip
\leftline{\bf DESCRIPTION OF THE DYC CIRCUIT}

Consult the FERA BUS TO VDAS INTERFACE schematic in Figures 2 and 3. 
VDAS [7] is the original name for an Event Fifo Buffer (EFB).
Trace the following major elements:

The primary data path:

INPUT DATA PORT (J1) to the

DIFFERENTIAL ECL TO TTL CONVERTERS (U3, U4, U6, U7) to the

DATA FIFO BUFFERS (U16, U17, U18, U19) to the

TTL TO RS-485 CONVERTERS (U21, U22 . . . U28) to the

OUTPUT DATA PORT (J4)

\bigskip
The ESN data path:

ESN INPUT PORT (J2) to the

RS-485 TO TTL CONVERTER (U13) to the

ESN LATCH REGISTER (U12) to the 

TTL TO RS-485 CONVERTER (U23) to the

OUTPUT DATA PORT (J4)

\bigskip
The word count data path:

INPUT DATA STROBE (DSTRB) to the

NIM TO TTL CONVERTER (U2) becoming the FERADS signal, to the

\hangindent=4mm \hangafter=1
WORD COUNT PAL (U20), where FERADS increments the count for each input
 data word, to the

TTL TO RS-485 CONVERTERS (U21, U22, U24) to the

OUTPUT DATA PORT (J4)

\begin{figure}[t!]
\centerline{\epsfxsize 5.6 truein \epsfbox{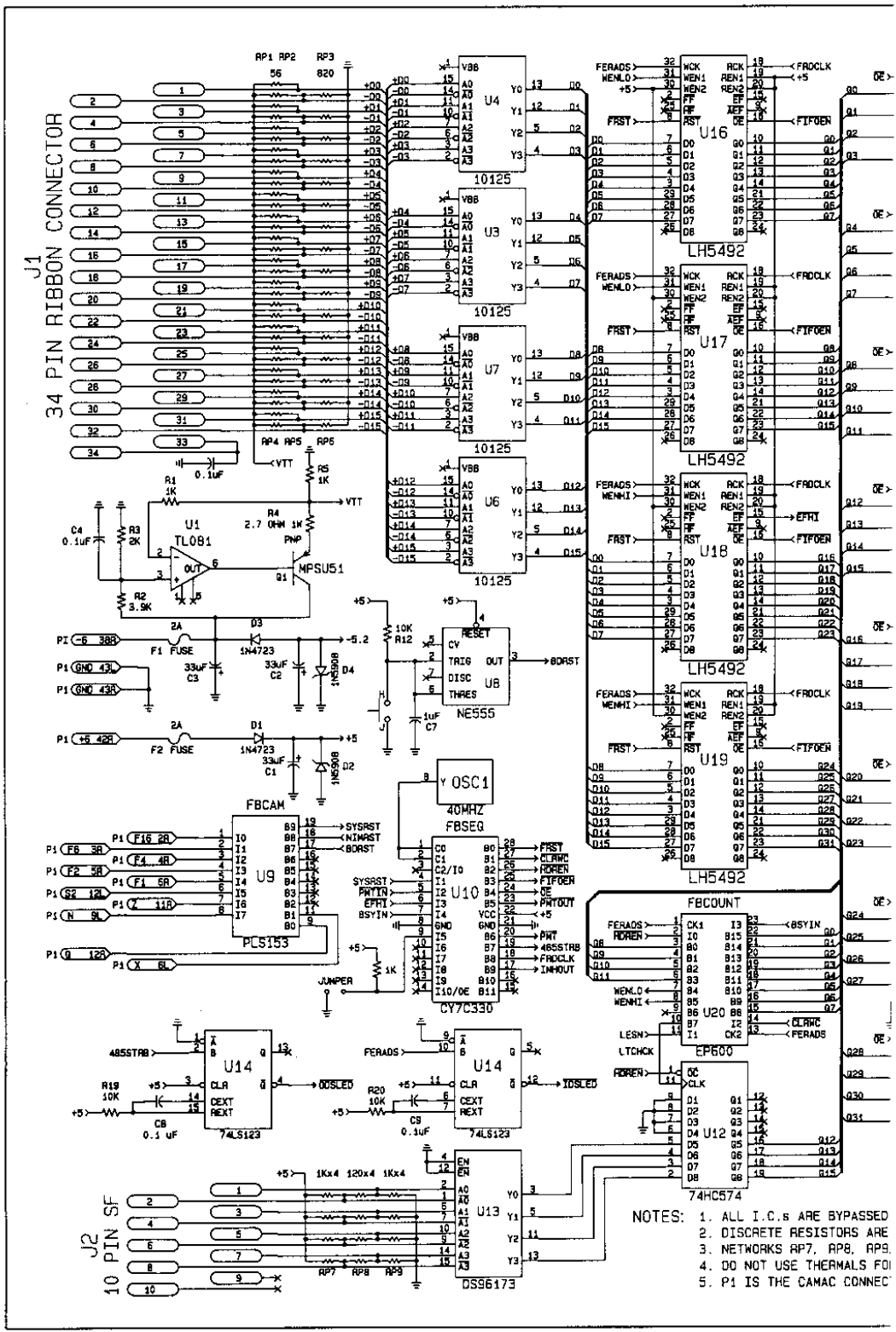}}
\vspace*{-2pt}
\caption[]{FERA BUS TO VDAS\,(EFB) INTERFACE Schematic (left side).}
\label{schematic_l}
\end{figure}

\begin{figure}[t!]
\centerline{\epsfxsize 5.7 truein \epsfbox{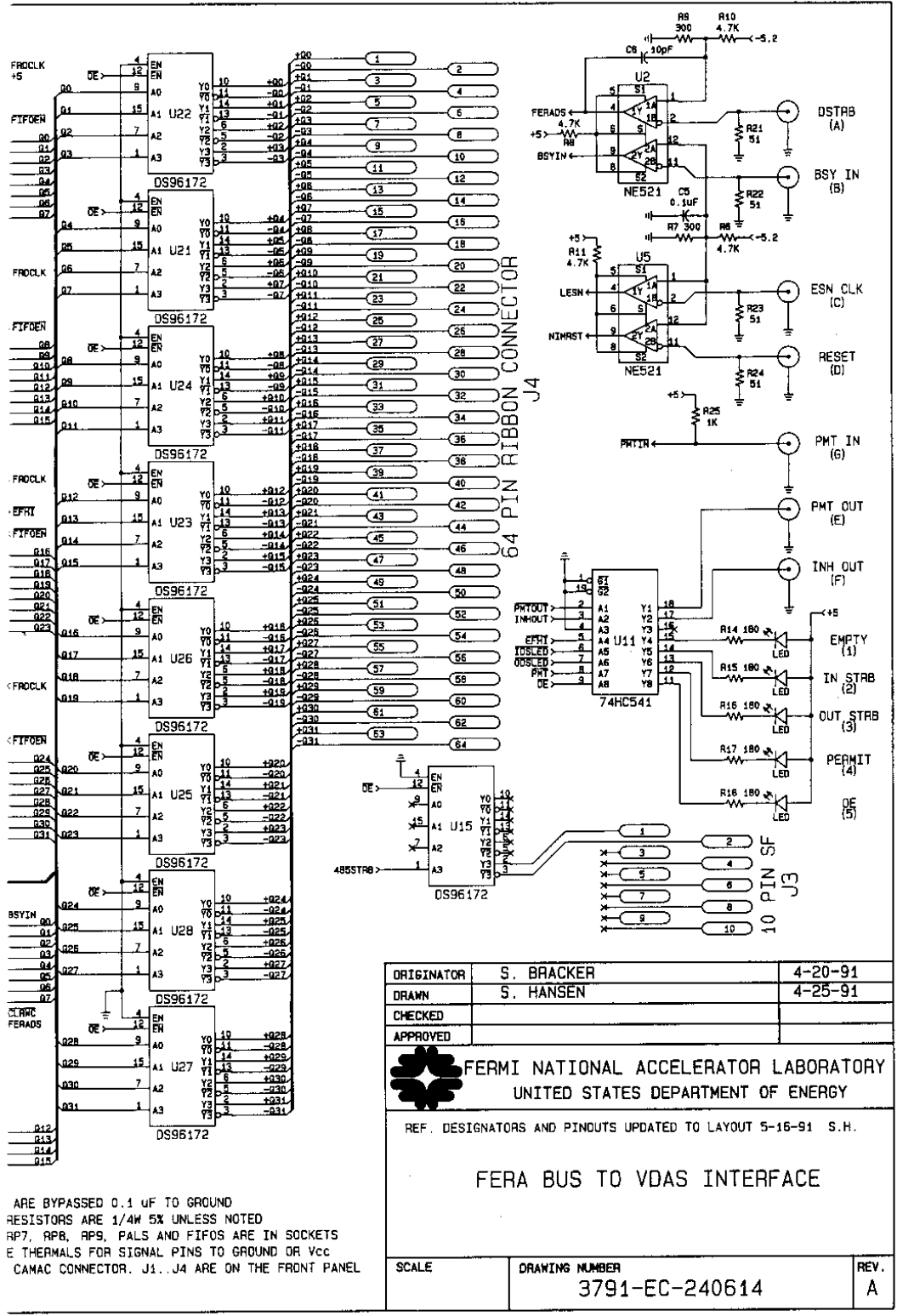}}
\vspace*{-2pt}
\caption[]{FERA BUS TO VDAS\,(EFB) INTERFACE Schematic (right side).}
\label{schematic_r}
\end{figure}

\bigskip
The camac and reset PAL (U9) accepts:

CAMAC FUNCTION LINES (F1, F2, F4, F8, F16)

THIS{\_}MODULE{\_}SELECTED LINE (N)

CRATE INITIALIZE LINE (Z)

CAMAC STROBE 2 (S2)

RESET SIGNAL FROM THE FRONT PANEL RESET LEMO (NIMRST)

POWER UP AND FRONT PANEL PUSHBUTTON RESET SIGNAL (BDRST)

\bigskip
The camac and reset PAL produces:

\hangindent=4mm \hangafter=1
SYSTEM RESET SIGNAL (SYSRST), which is asserted when a reset is required
 by camac command, front panel signal, reset pushbutton, or power up.

\hangindent=4mm \hangafter=1
CAMAC X AND Q SIGNALS, which are asserted whenever the module receives
 a legal camac command. The only legal camac command is reset: F9 . Ax . N.

\bigskip
The word count, fifo write enable and ESN latch strobe PAL (U20) accepts:

\hangindent=4mm \hangafter=1
CLEAR WORD COUNT signal (CLRWC) which zeros the word count upon reset and
 upon completion of processing for each event

\hangindent=4mm \hangafter=1
INPUT DATA STROBE (FERADS) which increments the word count for each input
 word received

\hangindent=4mm \hangafter=1
BUSY IN (BSYIN) which makes sure that no data is written to the data fifo 
 buffers unless the DYC is in its input phase, and may be used to clock the
 ESN latch

\hangindent=4mm \hangafter=1
LATCH ESN from the front panel connector (LESN) which may be used to clock 
 the ESN latch instead of BYSIN.

\bigskip
The word count, fifo write enable and ESN latch strobe PAL (U20) produces:

\hangindent=4mm \hangafter=1
WORD COUNT OUTPUTS (Q0, Q1 . . . Q11) which transmit the word count to the
 data output port when the header word is output

\hangindent=4mm \hangafter=1
FIFO WRITE ENABLE LINES (WENLO and WENHI) which enable input data to be
 clocked into the lower or upper half of the data fifo buffer, depending
 on the word count

\hangindent=4mm \hangafter=1
ESN LATCH CLOCK, the leading edge of which causes the ESN to be latched
 into the ESN latch register.

\bigskip
The master sequencer PAL (U10) is the state machine which controls the 
overall operation of the DYC. It accepts:

SYSTEM RESET (SYSRST) which is pulsed high upon receipt of any reset

\hangindent=4mm \hangafter=1
PERMIT IN (/PMTIN) which is pulsed low when the DYC receives the
 permit token from another module with which it shares the output bus.

\hangindent=4mm \hangafter=1
FIFO EMPTY FLAG (/EFHI) which is low when the data fifo buffer is 
 completely empty

\hangindent=4mm \hangafter=1
BUSY IN (BSYIN) which is asserted by the digitizer when data is about to
 be sent to the DYC, and negated by the digitizer when all data from the 
 event has arrived at the DYC

\hangindent=4mm \hangafter=1
FIRST MODULE JUMPER (FIRST). The jumper is inserted (FIRST is low) when
 this module is to be the first on the output bus to transmit its data.
 The jumper is removed (FIRST is high) when this module must await PERMIT
 IN to transmit its first event after a reset.

\bigskip
The master sequencer produces most of the signals which control the DYC:

\hangindent=4mm \hangafter=1
INHIBIT OUT (/INHOUT) is raised as soon as the DYC receives BUSY IN; it
 forbids the digitizing module to send another BUSY IN until INHOUT is
 negated. The DYC negates INHOUT after the current event is completely
 processed, and the DYC is ready to accept another event.

\hangindent=4mm \hangafter=1
FIFO READ CLOCK (FRDCLK) strobes the data fifo buffers, forcing them to
 deliver the next word of data to be output.

\hangindent=4mm \hangafter=1
OUTPUT DATA STROBE (485STRB) transmits a strobe pulse through the OUTPUT
 STROBE when the data in the OUTPUT DATA PORT is valid.

\hangindent=4mm \hangafter=1
PERMIT FLAG (/PMT) is low if the DYC has the permit token which allows it
 to drive the output bus, high if it does not have control of the bus until
 another PERMIT IN arrives.

\hangindent=4mm \hangafter=1
PERMIT OUT (/PMTOUT) transmits a short downgoing pulse through the PERMIT OUT
 front panel connector to transfer the permit token to the next module to
 deliver data on the output bus.

\hangindent=4mm \hangafter=1
OUTPUT ENABLE (/OE) tri-state enables the OUTPUT DATA PORT and OUTPUT STROBE
 RS-485 drivers. The drivers are enabled when /OE is low. The drivers must 
 not be enabled unless the DYC has the permit token.

\hangindent=4mm \hangafter=1
FIFO ENABLE OUTPUT (/FIFOEN) allows the data fifo buffer to transfer a word
 data from its memory to the OUTPUT DATA PORT. 

\hangindent=4mm \hangafter=1
HEADER ENABLE OUT (/HDREN) allows the word count PAL and the ESN latch to
 transfer the header word to the OUTPUT DATA PORT. 

\hangindent=4mm \hangafter=1
CLEAR WORD COUNT (/CLRWC) instructs the word count register to clear itself
 to zero during resets and at the end of each event.

\hangindent=4mm \hangafter=1
FIFO RESET (/FRST) instructs the data fifo buffers to clear themselves
 during resets and at the end of each event.

\bigskip
\bigskip
\bigskip
\leftline{\bf DESCRIPTION OF THE DYC SEQUENCER PAL}

Consult the PAL program listing in Appendix B for the DYC sequencer. 
It is divided into
several sections:

In the first section, signal names are assigned to pins and internal PAL
registers, and aliases for certain signal names are defined.

In the second section, all of the states of the state machine are defined
and given a state number. There are sixteen states defined; the DYC cycles
though 15 of them (all but RESET) each time it processes an event.

In the third section, the rules for changing state are given. Each 25 ns 
clock, the sequencer moves to a new state. The new state is dependent
on the old state, the current value of the input lines, and the current
value of the output lines.

In the fourth section, the values of the output lines are defined in terms
of the state and the value of the input lines.

In the fifth section, a long list of test vectors is given. The test vectors
specify a list of state changes that the sequencer should make; each vector
specifies the output list as a function of state and input list. The PAL
compiler checks to ensure that the PAL program as written (sections 1-4)
will produce the results specified. An error message is output if that
is not true.

\bigskip
\bigskip
\bigskip
\leftline{\bf NOTES ON THE DYC SEQUENCER STATES}

RESET: In response to any reset command, zero the word count, initialize the
data fifo buffer, and initialize the state machine. Then go wait for a BUSY IN.

WAITBSYHI: Wait for BUSY IN to be asserted, signaling that data from a new
event is about to begin. When BUSY IN does occur, advance to...

WAITSEND: Start the input phase of the DYC. Wait for BUSY IN to be negated,
signaling that all data from this event has arrived. During the WAITSEND
state, any data to be delivered by the digitizing module arrives at the
INPUT DATA PORT accompanied by INPUT STROBES (DSTRB, FERADS) which clock the
data into one section of the data fifo buffer. The timing of the incoming
data is asynchronous with respect to the DYC clock. 

HDRSTRBLO0: Start the output phase of the DYC. HDSTRBLO0 is the first of
four states which control the output of the header longword. HDRSTRBLO0
lasts 50 ns (two clock cycles), and then advances to...

HDRSTRBLO1: The second state controlling output of the header longword. 
It lasts 25 ns (one clock cycle), then advances immediately to...

HDRSTRBHI0: The third state controlling output of the header longword.
It lasts 25 ns (one clock cycle), then advances immediately to...

HDRSTRBHI1: The fourth state controlling output of the header longword.
It lasts 25 ns (one clock cycle), then advances immediately to...

OEGAP: Pause to ensure that the header data is no longer being sent to the
drivers of the OUTPUT DATA PORT. After this short delay, check to see if 
there is any data in the fifo. If there is, go to FIFOCLKHI0 to start 
transmitting data through the OUTPUT DATA PORT. If there is no data, skip 
ahead to PERMITGAP to complete processing of the event.

FIFOCLKHI0: The first state of four that transfer one longword of data from 
the fifo data buffer to the OUTPUT DATA PORT. This state and the three that 
follow are repeated for each longword to be transferred until the fifo 
buffers are completely empty. Advance immediately to...

FIFOCLKHI1: The second state of four that transfer a longword from the fifo 
data buffer to the OUTPUT DATA PORT. Advance immediately to...

FIFOCLKLO1: The third state of four that transfer a longword from the fifo 
data buffer to the OUTPUT DATA PORT. Advance immediately to...

FIFOCLKLO1: The last state of four that transfer a longword from the fifo 
data buffer to the OUTPUT DATA PORT. If the fifo is not empty yet, return to 
FIFOCLKHI0 to transfer the next longword. If the fifo is empty, then data 
output is complete; advance to...

PERMITGAP: A pause to ensure that the output drivers of the OUTPUT DATA 
PORT are disabled before a PERMIT OUT pulse is generated to pass control to 
another device sharing the output bus. Advance to:

PMTOUTLO0: The first of two states that reset the fifo buffers, zero the 
word count register, and transmit a PERMIT OUT pulse. Advance immediately 
to...

PMTOUTLO1: The second of two states that reset the fifo buffers, zero the 
word count register, and transmit a PERMIT OUT pulse. Advance immediately 
to...

INHIBITGAP: A brief pause to ensure that the DYC is ready to accept another 
event; then the INHIBIT OUT signal is negated. The digitizer may now 
transmit another event. Return to WAITBSYHI for the assertion of BUSY IN 
that signals the beginning of the next event....

Any RESET command will force the DYC to enter the RESET state; any data 
transfers in progress will be lost. Aside from this, the DYC always proceeds 
through the states in the prescribed order, looping only to transmit all the
words in the fifo buffer.

A few points to note in the state flow diagram:

By default, each state lasts 25 nanoseconds. In a few cases, states last 
until specified input conditions are true. In other cases, states need to last 
50 nanoseconds (two cycles). This is done by choosing a data output 
altered by the state, and advancing to the next state only after the data 
line has assumed its new value. All this could be avoided by defining a few 
more states.

At WAITSEND, the state machine cannot advance until BUSY IN has been negated 
and until the PERMIT token is present. These are the prerequisite 
conditions for starting to transmit through the OUTPUT DATA PORT.

At HDRSTRBLO0, if header enable is not asserted, then remain in this state 
for one more cycle. Header enable will be asserted during the first cycle; 
hence control will pass to the next state after the second HDRSTRBLO0 cycle.

At OEGAP, if BUSY IN has been negated and there is still no data in the 
fifo, then this is an empty event. The header has already been transmitted, 
but there is no data in the fifo, so go immediately to cleanup processing at 
PERMITGAP. In the normal case, when there is data present, go to FIFOCLKHI0 
to start transmitting it.

At FIFOCLKLO1, if the fifo buffer is empty and the output strobe is
asserted, then remain in this state for one more cycle. The output strobe
(/485STRB) will be negated during the first cycle; hence control will pass
to the next state after the second FIFOCLKLO1 cycle. If the fifo is not
empty, go back to transfer another longword at FIFOCLKHI0. 

At PERMITGAP, if output enable (to the OUTPUT DATA PORT drivers) is
asserted, then stay in this state for one more cycle. Output enable will be
negated in the first cycle; hence control will pass to the next state after
the second PERMITGAP cycle.

\eject
\leftline{\bf NOTES ON THE OUTPUT LINE DEFINITIONS:}

/FRST: Send a reset pulse to the fifo buffers during a reset, or for 50 ns
during the PMTOUTLO states near the end of processing for each event. 

/CLRWC: Send a pulse to zero the word counter during a reset, or for 50 ns
during the PMTOUTLO states near the end of processing for each event. At
this time, /FRST and /CLRWC adhere to the same timing.

/HDREN: Gate the header information (word count and ESN) onto the output 
drivers for the OUTPUT DATA PORT during all four HDRSTRB cycles near the 
start of the output phase. The header should be enabled for a total of 125 
nanoseconds.

/FIFOEN: Gate the data words from the data fifo buffer onto the output
drivers for the OUTPUT DATA PORT during all four FIFOCLK cycles. These four 
cycles are repeated (and /FIFOEN is held) until the fifo buffer is empty.
Fifo enable for each longword output lasts 100 nanoseconds.

/PMTOUT: Transmit a PERMIT OUT pulse to the next module to use the output 
bus. The pulse is 50 ns long.

/PMT: This is the "I have the permit token" flipflop. Set the flipflop (/PMT
is low) if a PERMIT IN pulse arrives or upon RESET if the FIRST jumper is
installed. Clear the flipflop (/PMT is high) at PERMITGAP (immediately after
the last data word is transmitted) or upon RESET if the FIRST jumper is
removed. 

/OE: This signal is low if the RS-485 output drivers are tri-state enabled. 
It has the same timing as /PMT. If the permit token is present, then the 
output lines are driven, even if the DYC is not presently transmitting 
anything. If the permit token is absent, it is illegal for the DYC to drive 
the output lines.

/485STRB: Transmit a 50 ns OUTPUT STROBE for the header longword and for 
every data longword.

FRDCLK: On the rising edge of this signal, clock out the next longword of
data from the fifo. If this is the last longword in the fifo, the /EFHI line 
(empty fifo) line is also asserted (set low). FRDCLK and /485STRB are so 
related in time that the asserted (rising, trailing) edge of the OUTPUT DATA 
STROBE occurs about in the middle of a 70 ns data validity interval on the 
OUTPUT DATA PORT.

INHOUT: Arrival of the next event is forestalled until the DYC is done 
processing the current event. INHOUT is asserted immediately following the 
assertion of a BUSY IN, and remains asserted until the very end of event 
processing, when the DYC returns to the WAITBSYHI state.

\bigskip
\bigskip
\bigskip
\leftline{\bf COMMENTARY ON LOGIC ANALYZER SCREENS [14]}

SCREEN 1: The DYC is reset. The fifo and the word count are marked empty
(/CLRWC and /FRST). The FIRST jumper is removed (FIRST is high) so the /PMT
flag is high (token not present). 

The digitizer asserts BUSY IN. The input phase begins. The DYC immediately
asserts INHIBIT OUT (INHOUT) to make sure that another event is forestalled
until the DYC is ready to receive it. Data from the current event are not
inhibited. Twelve words of data arrive at the INPUT DATA PORT (not shown).
While data is still arriving, the PERMIT IN signal arrives; the DYC
immediately sets its /PMT flag low and enables the output drivers. No output
is produced until all the input data has arrived; this is signified by the
negation of BUSY IN. 

The negation of BUSY IN begins the output phase. The header information 
(word count and ESN) are gated to the OUTPUT DATA PORT, and the OUTPUT 
STROBE is pulsed (first 485STB). The header information is disabled and the 
fifo data (one longword at a time) is gated to the OUTPUT DATA PORT by
/FIFOE. Output words are clocked from the fifo by FRDCLK. When the data is 
valid, OUTPUT STROBEs are produced (all but the first 485STB). When the fifo 
is empty (/EFHI is low), data output ends. The output drivers are disabled, 
and soon thereafter the word count is cleared, the fifos are cleared, and a 
PERMIT OUT pulse is set to hand over the output bus to the next device to 
use it. When all of this has been done, and the DYC is ready to receive a 
new event, INHIBIT OUT is removed. The DYC now idles, waiting for the next 
event.

SCREEN 2: The same as SCREEN 1, except that PERMIT IN arrives before the 
input phase begins. This is fine; the DYC accepts the permit token, turns on 
its output drivers, and waits for the next event. All remaining processing 
proceeds as above. Eleven data words arrive at the INPUT DATA PORT (not 
shown), producing six longwords of output, preceded by the header longword.

SCREEN 3: The same as SCREEN 1, except that PERMIT IN is delayed until after 
all of the data from this event has arrived, and BUSY IN has been negated. 
Output is delayed until the permit token arrives; it then proceeds as above.
Eighteen data words arrive at the INPUT DATA PORT (not shown), producing 
nine longwords of output, preceded by the header longword.

SCREEN 4: The same as SCREEN 1, except that no data arrives from the 
digitizer (/EFHI stays low). The permit token arrives, and later the negated 
edge of BUSY IN marks the end of data transmission; none has arrived. The 
header is output, producing a single data strobe. No data output cycles are 
performed. The cleanup phase proceeds as above.

\bigskip
\bigskip
\bigskip
\leftline{\bf USER'S GUIDE TO THE DYC FRONT PANEL}

The DYC front panel has seven LEMO connectors, four ribbon cable connectors,
five lights, and one pushbutton.

RESET pushbutton \hfill \break 
Pushing the reset button aborts any operation in progress and prepares the 
DYC to process the next event.

EMPTY light (yellow) \hfill \break
On when the fifo data buffer is empty. During the spill, the empty light 
will be partially on. If the data rate is very high, then the fifo will 
usually have data in it, and the light will be dim. At very low data rates, 
the light may blink perceptibly. During the interspill, the empty light 
should always be on.

IN STROBE light (red) \hfill \break
Pulsed on whenever a data word arrives at the INPUT DATA PORT. During the 
spill, the light will be lit whenever input data arrives. At very low data 
rates, the light may blink perceptibly. During the interspill, the IN STROBE 
light should always be off.

OUT STROBE light (green) \hfill \break
Pulsed on whenever a data word is transmitted from the OUTPUT DATA PORT. 
During the spill, the light will be lit whenever output data is being sent 
to the Event Fifo Buffer. At very low data rates, the light may blink 
perceptibly. During the interspill, the OUT STROBE light should always be 
off.

PERMIT light (yellow) \hfill \break
On whenever the module has the PERMIT TOKEN, and is thereby authorized to 
drive the output bus. After a reset (and before any other control signals 
have been asserted), the PERMIT light should be on if the FIRST jumper is 
installed, and off if the FIRST jumper is removed. (The FIRST jumper is 
installed if this DYC is the first device to transmit data for each event.)
The PERMIT light will always be on during data transmission, though at very 
low data rates it may be too faint to see clearly. The light will be 
turned out when the PERMIT TOKEN is passed to another module, and turned 
back on when a PERMIT IN arrives, passing the token back. During the 
interspill, the PERMIT light should be on if the FIRST jumper is installed, 
otherwise off.

OUTPUT ENABLE light (yellow) \hfill \break
At the present time, this light follows the PERMIT light exactly; refer to 
the paragraph above for a description of its operation.

DSTRB input connector \hfill \break
Receives a short NIM pulse for each data word received at the INPUT DATA PORT.

BSY IN input connector \hfill \break
Receives a long NIM pulse which is asserted before each event's data starts 
arriving, and negated after the event's data has all been transmitted.

ESN CLK input connector \hfill \break
Not used in E--791; can receive a short NIM pulse to latch the data at the 
ESN PORT into the DYC. In E--791, the ESN is latched at the asserted edge of 
BSY IN.

RESET input connector \hfill \break
Receives a NIM pulse which will reset the DYC (same as the pushbutton).

PMT IN input connector \hfill \break
Receives a short active low TTL pulse which moves the PERMIT TOKEN to the 
DYC from the module that previously had it.

PMT OUT output connector

INH OUT output connector

\bigskip
\bigskip
\bigskip
\leftline{\bf ACKNOWLEDGEMENTS}

Many thanks to Dan Kleinert, George Wolf, and Stephen Pordes for 
their roles in this endeavor.
Also a special thanks to Barry Lasker for an introduction to data
acquisition [15]. 
This work was supported by the U.~S.~Department of Energy
(DE-AC02-76CHO3000).

\bigskip
\bigskip
\bigskip
\leftline{\bf REFERENCES}


\begin{enumerate}

\item http://www.lecroy.com/lrs/dsheets/4300b.htm

\item D.~Bartlett et al., Nucl.~Instrum.~Meth. {\bf A260} (1987) 55.

\item V.~K.~Bharadwaj et al., Nucl.~Instrum.~Meth. {\bf 155} (1978) 411; 
{\bf A228} (1985) 283; \hfill \break
D.~J.~Summers, Nucl.~Instrum.~Meth. {\bf A228} (1985) 290.

\item J.~A. Appel et al.,  Nucl.~Instrum.~Meth. {\bf A243} (1986) 361.

\item S.~Amato et al., {\it The E791 Parallel Architecture Data Acquisition 
System,} Nucl.~Instrum.~Meth. {\bf A324} (1993) 535.

\item Steve Bracker et al. (Fermilab E--791), {\it A Simple Multiprocessor
Management System for Event Parallel Computing,}
IEEE Trans.~Nucl.~Sci. {\bf 43} (1996) 2457.

\item A.~E.~Baumbaugh et al., 
IEEE Trans.~Nucl.~Sci. {\bf 33} (1986) 903; \hfill\break
K.~L.~Knickerbocker et al., IEEE Trans.~Nucl.~Sci. {\bf 34} (1987) 245.

\item S.~Hansen et al., 
IEEE Trans.~Nucl.~Sci. {\bf 34} (1987) 1003; \hfill\break
M.~Bernett et al., IEEE Trans.~Nucl.~Sci. {\bf 34} (1987) 1047; \hfill\break
R.~Vignoni et al., IEEE Trans.~Nucl.~Sci. {\bf 34} (1987) 756; \hfill\break
C.~Gay and S.~Bracker, IEEE Trans.~Nucl.~Sci. {\bf 34} (1987) 870.

\item E.~M.~Aitala et al. (Fermilab E--791), {\em Experimental Evidence for a
Light and Broad Scalar Resonance in $D^+ \rightarrow \pi^- \pi^+ \pi^+$ Decay,}
Phys.~Rev.~Lett. {\bf 86} (2001) 770; \hfill\break
\vskip -13pt
E.~M.~Aitala et al. (Fermilab E--791), {\it Branching Fractions for $D^0
\rightarrow K^+K^-$ and $D^0 \rightarrow \pi^+\pi^-$, and a Search for CP
Violation in $D^0$ Decays,} Phys.~Lett. {\bf B421} (1998) 405.

\item M.~Ambrogiani et al (Fermilab E--835), {\it Study of the 
$\chi_{c\,0} (1^3 P_0)$ State of Charmonium Formed in $\overline{p} p$
Annihilations,} Phys.~Rev.~Lett. {\bf 83} (1999) 2902; \hfill\break
\vskip -13pt
M.~Ambrogiani et al (Fermilab E--835), {\it Study of the Angular Distributions
of the Reactions $\overline{p} p \rightarrow \chi_{c\,1} , \, \chi_{c\,2}
\rightarrow J/\psi \,\, \gamma \rightarrow e^+ e^- \gamma$,} Phys.~Rev. 
{\bf D65} (2002) 052002.

\item J.~M.~Link et al. (Fermilab FOCUS), {\it A Measurement of the Lifetime
Differences in the Neutral D Meson System,} Phys.~Lett. {\bf B485} (2000) 62;
\hfill\break
\vskip -13pt
J.~M.~Link et al. (Fermilab FOCUS), {\it Search for CP Violation in $D^0$ and
$D^+$ Decays,} Phys.~Lett. {\bf B491} (2000) 232.

\item S.~Y.~Jun et al. (Fermilab SELEX), {\it Observation of the Cabibbo 
Suppressed Decay $\Xi_c^+ \rightarrow p K^- \pi^+$,} 
Phys.~Rev.~Lett. {\bf 84} (2000) 1857; \hfill\break
\vskip -13pt
A.~Kushnirenko et al. (Fermilab SELEX)
{\it Precision Measurements of the $\Lambda_c^+$ and $D^0$ Lifetimes,} 
Phys.~Rev.~Lett. {\bf 86} (2001) 5243.

\item A.~Alavi--Harati et al. (Fermilab KTeV), {\it Observation of Direct CP
Violation in $K_{S,L} \rightarrow \pi \pi$ Decays,}
Phys.~Rev.~Lett. {\bf 83} (1999) 22; \hfill\break
\vskip -13pt
A.~Alavi--Harati et al. (Fermilab KTeV),
{\it Measurement of the Branching Ratio and Form Factor of 
$K_L \rightarrow \mu^+ \mu^- \gamma$,}
Phys.~Rev.~Lett. {\bf 87} (2001) 111802. 

\item Hewlett Packard 1651A Logic Analyzer, 100 MHz, 32 channels.

\item Barry M. Lasker, Stephen B. Bracker, and William E. Kunkel,
Publ.~Astron.~Soc.~Pac. {\bf 85} (1973) 109.  

\end{enumerate}

\eject
\leftline{\bf APPENDIX A}
\leftline{\bf TESTS MADE TO VERIFY THE DYC DESIGN}

\begin{enumerate}
\item PERMIT OUT from the SCC was checked. It is normally high, and lowered for 
about 300 ns when the token is passed.

\item INHIBIT OUT from a working DYC was checked. It is normally low, and 
raised for 20-40 microseconds while the DYC is busy processing an event.

\item The time from assertion of BUSY IN to the first data strobe was measured
for all three systems; it is at least 300 ns for all systems.

\item The time from the last data strobe to the negation of BUSY IN was 
measured for all systems; it is at least 64 ns for all systems.
\end{enumerate}

\bigskip
\bigskip
\bigskip
\leftline{\bf APPENDIX B}
\leftline{\bf DYC SEQUENCER PAL PROGRAM LISTING: FBSEQ.ABL}
\begin{verbatim}
Module Sequencer              flag '-r4'

title 'Controls transmission of FIFO data in response to Busy In.
       Responsible for receiving Permit in and Sending Permit Out
       and Inhibit Out.

       Sten Hansen             Fermilab               5-7-91'

       FbSeq                  device                 'P330';

"Inputs:                        location                 U10
Clk1     "Macro Cell clock"      pin  1;
Clk2     "Input register clock"  pin  2;

SysRst,PmtIn,EFHi                pin  4,5,6;
BsyIn,First                      pin  7,9;

" Outputs:
FRst,ClrWC,HdrEn,FifoEn,OE       pin  28,27,26,25,24;
PmtOut,Pmt,_485Strb              pin  23,20,19;
FRDClk,InhOut                    pin  18,17;

"Buried state registers
S3,S2,S1,S0                    node 34,33,32,31;

"Shorten Signal names to allow 1 page width test vectors
BI,PI,EF,FR,CW,HEn,FEn,PO,TS,FC,InhO,Rst = BsyIn,PmtIn,EFHi,FRst,ClrWC,
                        HdrEn,FifoEn,PmtOut,_485Strb,FRDClk,InhOut,SysRst;

"Simplify don't care and clock terms..
C,X  = .C.,.X.;

"Group related signals into sets..
Mode = [S3,S2,S1,S0];

"Avoid any false transitions by making only 1 bit state changes
"in response to external asynchronous signals..

Reset      = ^b0000;
WaitBsyHi  = ^b0001;
WaitSend   = ^b0101;
HdrStrbLo0 = ^b0100;
HdrStrbLo1 = ^b0110;
HdrStrbHi0 = ^b0111;
HdrStrbHi1 = ^b1111;
OEGap      = ^b1110;
FifoClkHi0 = ^b1100;
FifoClkHi1 = ^b1101;
FifoClkLo0 = ^b1001;
FifoClkLo1 = ^b1000;
PermitGap  = ^b1010;
PmtOutLo0  = ^b1011;
PmtOutLo1  = ^b0011;
InhibitGap = ^b0010;

@if 0
{
Note:
     There are 2 feedback paths on the IO pins. Feedback from the pin goes
     through an input register. Direct feedback from the !Q output of the
     output register is specified with an OutputName.Q extension. ABEL does
     not automatically invert the sense of the feedback however. Therefore
     all references to a .Q output are really references to a .!Q output and
     must be used as an active low signal, regardless of the polarity of the
     output.
}

state_diagram Mode

 state      Reset: if !SysRst then WaitBsyHi else Reset;

 state  WaitBsyHi: if BsyIn & !SysRst then WaitSend
                   else if SysRst then Reset
                   else WaitBsyHi;

 state   WaitSend: if !BsyIn & !SysRst & Pmt.Q then HdrStrbLo0
                   else if SysRst then Reset
                   else WaitSend;

 state HdrStrbLo0: if !SysRst & HdrEn.Q then HdrStrbLo1
                   else if SysRst then Reset
                   else HdrStrbLo0;
 state HdrStrbLo1: if !SysRst then HdrStrbHi0 else Reset;

 state HdrStrbHi0: if !SysRst then HdrStrbHi1 else Reset;
 state HdrStrbHi1: if !SysRst then OEGap      else Reset;

 state      OEGap: if !EFHi & !SysRst then PermitGap
                   else if SysRst then Reset
                   else FifoClkHi0;

 state FifoClkHi0: if !SysRst then FifoClkHi1 else Reset;
 state FifoClkHi1: if !SysRst then FifoClkLo0 else Reset;
 state FifoClkLo0: if !SysRst then FifoClkLo1 else Reset;

 state FifoClkLo1: if !EFHi & !_485Strb.Q & !SysRst then PermitGap
                   else if !EFHi & _485Strb.Q then FifoClkLo1
                   else if SysRst then Reset
                   else FifoClkHi0;

 state  PermitGap: if !SysRst & OE then PmtOutLo0
                   else if SysRst then Reset
                   else PermitGap;

 state  PmtOutLo0: if !SysRst then PmtOutLo1  else Reset;
 state  PmtOutLo1: if !SysRst then InhibitGap else Reset;

 state InhibitGap: if !SysRst then WaitBsyHi  else Reset;

equations

"For now, there is no difference between Fifo Reset and Clear Word Count
!FRst    := (Mode == Reset) # (Mode == PmtOutLo0) # (Mode == PmtOutLo1);

!ClrWC   := (Mode == Reset) # (Mode == PmtOutLo0) # (Mode == PmtOutLo1);

!HdrEn   := (Mode == HdrStrbLo0) # (Mode == HdrStrbLo1)
          # (Mode == HdrStrbHi0) # (Mode == HdrStrbHi1);

!FifoEn  := (Mode == FifoClkHi0) # (Mode == FifoClkHi1)
          # (Mode == FifoClkLo0) # (Mode == FifoClkLo1);

!PmtOut  := (Mode == PmtOutLo0) # (Mode == PmtOutLo1);

"If the jumper is in  (First low)    set Pmt on Reset
"If the jumper is out (First High) clear Pmt on Reset
!Pmt     := !PmtIn # SysRst & !First
        # Pmt.Q & !((Mode == PermitGap) # SysRst & First);

"For now RS-485 OE is the same as Pmt..
!OE      := !PmtIn # SysRst & !First
        # Pmt.Q & !((Mode == PermitGap) # SysRst & First);

"Baumbaugh buffers clock on the rising edge, but need a default state of high
"in order to reset properly..
!_485Strb := (Mode == HdrStrbLo0) & HdrEn.Q # (Mode == HdrStrbLo1)
           # (Mode == FifoClkHi1) # (Mode == FifoClkLo0);

FRDClk    := (Mode == FifoClkHi0) # (Mode == FifoClkHi1);

InhOut    := BsyIn # !InhOut.Q & !(Mode == WaitBsyHi);

test_vectors
([Clk1,Clk2,Rst,BI,PI,EF,First]->[Mode,CW,HEn,FEn,PO,Pmt,TS,FC,InhO]);
"Read sequence with 'First' jumper out..
[ C   , C  , 1 ,0 ,1 ,0 ,  1  ]->[ 1  ,0 ,1  , 1 ,1 , 0 ,1 ,0 , 1  ];
[ C   , C  , 0 ,0 ,1 ,0 ,  1  ]->[ 0  ,1 ,1  , 1 ,1 , 1 ,1 ,0 , 0  ];
[ C   , C  , 0 ,0 ,1 ,0 ,  1  ]->[ 1  ,0 ,1  , 1 ,1 , 1 ,1 ,0 , 0  ];
[ C   , C  , 0 ,0 ,1 ,0 ,  1  ]->[ 1  ,1 ,1  , 1 ,1 , 1 ,1 ,0 , 0  ];
[ C   , C  , 0 ,0 ,1 ,0 ,  1  ]->[ 1  ,1 ,1  , 1 ,1 , 1 ,1 ,0 , 0  ];"5
[ C   , C  , 0 ,1 ,1 ,0 ,  1  ]->[ 1  ,1 ,1  , 1 ,1 , 1 ,1 ,0 , 0  ];
[ C   , C  , 0 ,1 ,1 ,0 ,  1  ]->[ 5  ,1 ,1  , 1 ,1 , 1 ,1 ,0 , 1  ];
[ C   , C  , 0 ,1 ,1 ,1 ,  1  ]->[ 5  ,1 ,1  , 1 ,1 , 1 ,1 ,0 , 1  ];
[ C   , C  , 0 ,1 ,1 ,1 ,  1  ]->[ 5  ,1 ,1  , 1 ,1 , 1 ,1 ,0 , 1  ];
[ C   , C  , 0 ,1 ,1 ,1 ,  1  ]->[ 5  ,1 ,1  , 1 ,1 , 1 ,1 ,0 , 1  ];"10
[ C   , C  , 0 ,1 ,1 ,1 ,  1  ]->[ 5  ,1 ,1  , 1 ,1 , 1 ,1 ,0 , 1  ];
[ C   , C  , 0 ,1 ,1 ,1 ,  1  ]->[ 5  ,1 ,1  , 1 ,1 , 1 ,1 ,0 , 1  ];
[ C   , C  , 0 ,0 ,1 ,1 ,  1  ]->[ 5  ,1 ,1  , 1 ,1 , 1 ,1 ,0 , 1  ];
[ C   , C  , 0 ,0 ,1 ,1 ,  1  ]->[ 5  ,1 ,1  , 1 ,1 , 1 ,1 ,0 , 1  ];
[ C   , C  , 0 ,0 ,0 ,1 ,  1  ]->[ 5  ,1 ,1  , 1 ,1 , 1 ,1 ,0 , 1  ];"15
[ C   , C  , 0 ,0 ,1 ,1 ,  1  ]->[ 5  ,1 ,1  , 1 ,1 , 0 ,1 ,0 , 1  ];
[ C   , C  , 0 ,0 ,1 ,1 ,  1  ]->[ 4  ,1 ,1  , 1 ,1 , 0 ,1 ,0 , 1  ];
[ C   , C  , 0 ,0 ,1 ,1 ,  1  ]->[ 4  ,1 ,0  , 1 ,1 , 0 ,1 ,0 , 1  ];
[ C   , C  , 0 ,0 ,1 ,1 ,  1  ]->[ 6  ,1 ,0  , 1 ,1 , 0 ,0 ,0 , 1  ];
[ C   , C  , 0 ,0 ,1 ,1 ,  1  ]->[ 7  ,1 ,0  , 1 ,1 , 0 ,0 ,0 , 1  ];"20
"Clk1,Clk2,Rst,BI,PI,EF,First]->[Mode,FR,HEn,FEn,PO,Pmt,TStb,FClk,InhO]);
[ C   , C  , 0 ,0 ,1 ,1 ,  1  ]->[ 15 ,1 ,0  , 1 ,1 , 0 ,1 ,0 , 1  ];
[ C   , C  , 0 ,0 ,1 ,1 ,  1  ]->[ 14 ,1 ,0  , 1 ,1 , 0 ,1 ,0 , 1  ];
[ C   , C  , 0 ,0 ,1 ,1 ,  1  ]->[ 12 ,1 ,1  , 1 ,1 , 0 ,1 ,0 , 1  ];
[ C   , C  , 0 ,0 ,1 ,1 ,  1  ]->[ 13 ,1 ,1  , 0 ,1 , 0 ,1 ,1 , 1  ];
[ C   , C  , 0 ,0 ,1 ,1 ,  1  ]->[ 9  ,1 ,1  , 0 ,1 , 0 ,0 ,1 , 1  ];"25
[ C   , C  , 0 ,0 ,1 ,1 ,  1  ]->[ 8  ,1 ,1  , 0 ,1 , 0 ,0 ,0 , 1  ];
[ C   , C  , 0 ,0 ,1 ,1 ,  1  ]->[ 12 ,1 ,1  , 0 ,1 , 0 ,1 ,0 , 1  ];
[ C   , C  , 0 ,0 ,1 ,1 ,  1  ]->[ 13 ,1 ,1  , 0 ,1 , 0 ,1 ,1 , 1  ];
[ C   , C  , 0 ,0 ,1 ,1 ,  1  ]->[ 9  ,1 ,1  , 0 ,1 , 0 ,0 ,1 , 1  ];
[ C   , C  , 0 ,0 ,1 ,1 ,  1  ]->[ 8  ,1 ,1  , 0 ,1 , 0 ,0 ,0 , 1  ];"30
[ C   , C  , 0 ,0 ,1 ,1 ,  1  ]->[ 12 ,1 ,1  , 0 ,1 , 0 ,1 ,0 , 1  ];
[ C   , C  , 0 ,0 ,1 ,1 ,  1  ]->[ 13 ,1 ,1  , 0 ,1 , 0 ,1 ,1 , 1  ];
[ C   , C  , 0 ,0 ,1 ,0 ,  1  ]->[ 9  ,1 ,1  , 0 ,1 , 0 ,0 ,1 , 1  ];
[ C   , C  , 0 ,0 ,1 ,0 ,  1  ]->[ 8  ,1 ,1  , 0 ,1 , 0 ,0 ,0 , 1  ];
[ C   , C  , 0 ,0 ,1 ,0 ,  1  ]->[ 8  ,1 ,1  , 0 ,1 , 0 ,1 ,0 , 1  ];"35
[ C   , C  , 0 ,0 ,1 ,0 ,  1  ]->[ 10 ,1 ,1  , 0 ,1 , 0 ,1 ,0 , 1  ];
[ C   , C  , 0 ,0 ,1 ,0 ,  1  ]->[ 10 ,1 ,1  , 1 ,1 , 1 ,1 ,0 , 1  ];
[ C   , C  , 0 ,0 ,1 ,0 ,  1  ]->[ 11 ,1 ,1  , 1 ,1 , 1 ,1 ,0 , 1  ];
[ C   , C  , 0 ,0 ,1 ,0 ,  1  ]->[ 3  ,0 ,1  , 1 ,0 , 1 ,1 ,0 , 1  ];
[ C   , C  , 0 ,0 ,1 ,0 ,  1  ]->[ 2  ,0 ,1  , 1 ,0 , 1 ,1 ,0 , 1  ];"40
[ C   , C  , 0 ,0 ,1 ,0 ,  1  ]->[ 1  ,1 ,1  , 1 ,1 , 1 ,1 ,0 , 1  ];
[ C   , C  , 0 ,0 ,1 ,0 ,  1  ]->[ 1  ,1 ,1  , 1 ,1 , 1 ,1 ,0 , 0  ];
[ C   , C  , 0 ,0 ,1 ,0 ,  1  ]->[ 1  ,1 ,1  , 1 ,1 , 1 ,1 ,0 , 0  ];
[ C   , C  , 0 ,0 ,1 ,0 ,  0  ]->[ 1  ,1 ,1  , 1 ,1 , 1 ,1 ,0 , 0  ];
[ C   , C  , 0 ,0 ,1 ,0 ,  0  ]->[ 1  ,1 ,1  , 1 ,1 , 1 ,1 ,0 , 0  ];"45
"Clk1,Clk2,Rst,BI,PI,EF,First]->[Mode,FR,HEn,FEn,PO,Pmt,TStb,FClk,InhO]);

End  Sequencer
^Z
\end{verbatim}

\bigskip
\bigskip
\bigskip
\leftline{\bf APPENDIX C}
\leftline{\bf DYC WORD COUNT PAL PROGRAM LISTING: FBCOUNT.ABL}
\begin{verbatim}
Module  Address_Counter

Title   'Increments address counter with each Fera Data strobe
         Alternates Fifo enables with each strobe

        Sten Hansen       Fermilab Physics Dept.        5-3-91'

        FBCOUNT                device          'E0600';

       "Location                                 U20

        Clk1,ClkESN,Clk2            pin    1,10,13;
        HdrEn,ClrWC,BsyIn           pin    2,14,23;

        Q0,Q1,Q2,Q3,Q4,Q5           pin    22,21,20,19,18,17;
        Q6,Q7,Q8,Q9,Q10,Q11         pin    16,15,3,4,5,6;
        EnFLo,EnFHi                 pin    7,8;

N = 211;  "Set to the desired number of cycles run in the test vectors

Q0,Q1,Q2,Q3,Q4,Q5,Q6,Q7,Q8,Q9,Q10,Q11           IsType 'Reg_T,Feed_Reg';

Q0.RE,Q1.RE,Q2.RE,Q3.RE,Q4.RE,Q5.RE,Q6.RE,
Q7.RE,Q8.RE,Q9.RE,Q10.RE,Q11.RE                 IsType 'Eqn';

C,Z,X = .C.,.Z.,.X.;
Count = [Q11..Q0];


Equations

"For the time being clock the ESN latch with the leading edge of Busy In
 ClkESN  =  BsyIn;

"A 12 bit counter constructed from T FF's
"If the quantity on the right side of the equation is true, the FF toggles

 Q0 := BsyIn;
 Q1 :=                                                    Q0;
 Q2 :=                                               Q1 & Q0;
 Q3 :=                                          Q2 & Q1 & Q0;
 Q4 :=                                     Q3 & Q2 & Q1 & Q0;
 Q5 :=                                Q4 & Q3 & Q2 & Q1 & Q0;
 Q6 :=                           Q5 & Q4 & Q3 & Q2 & Q1 & Q0;
 Q7 :=                      Q6 & Q5 & Q4 & Q3 & Q2 & Q1 & Q0;
 Q8 :=                 Q7 & Q6 & Q5 & Q4 & Q3 & Q2 & Q1 & Q0;
 Q9 :=            Q8 & Q7 & Q6 & Q5 & Q4 & Q3 & Q2 & Q1 & Q0;
Q10 :=       Q9 & Q8 & Q7 & Q6 & Q5 & Q4 & Q3 & Q2 & Q1 & Q0;
Q11 := Q10 & Q9 & Q8 & Q7 & Q6 & Q5 & Q4 & Q3 & Q2 & Q1 & Q0;

"Clear all FFs to 0 with ClrWC
 [Q0.RE,Q1.RE,Q2.RE,Q3.RE,Q4.RE,Q5.RE,Q6.RE,Q7.RE,Q8.RE,
  Q9.RE,Q10.RE,Q11.RE] = !ClrWC;

"Turn tri-states on with !HdrEn
 [Q0.OE,Q1.OE,Q2.OE,Q3.OE,Q4.OE,Q5.OE,Q6.OE,Q7.OE,Q8.OE,
  Q9.OE,Q10.OE,Q11.OE] = !HdrEn;

EnFHi = !Q0 & ClrWC & BsyIn;
EnFLo =  Q0 & ClrWC & BsyIn;

Test_Vectors

([Clk1,Clk2,HdrEn,ClrWC,BsyIn]->[Count,EnFHi,EnFLo,ClkESN]);
 [ X  , X  ,  0  ,  0  ,  X  ]->[  0  ,   0 ,   0 ,  X];
 [ X  , X  ,  0  ,  1  ,  0  ]->[  0  ,   0 ,   0 ,  0];
 [ C  , C  ,  1  ,  1  ,  0  ]->[  Z  ,   0 ,   0 ,  0];
 [ C  , C  ,  1  ,  1  ,  0  ]->[  Z  ,   0 ,   0 ,  0];
 [ 0  , 0  ,  0  ,  1  ,  0  ]->[  0  ,   0 ,   0 ,  0];
@repeat N
{
 [ C  , C  ,  1  ,  1  ,  1  ]->[  Z  ,   0 ,   1 ,  1];
 [ C  , C  ,  1  ,  1  ,  1  ]->[  Z  ,   1 ,   0 ,  1];
}
 [ 0  , 0  ,  0  ,  1  ,  0  ]->[  2*N,   0 ,   0 ,  0];
 [ C  , C  ,  1  ,  1  ,  1  ]->[  Z  ,   0 ,   1 ,  1];
 [ 0  , 0  ,  0  ,  1  ,  0  ]->[  2*N+1, 0 ,   0 ,  0];
 [ 0  , 0  ,  0  ,  0  ,  0  ]->[  0  ,   0 ,   0 ,  0];

end Address_Counter

\end{verbatim}

\bigskip
\bigskip
\bigskip
\leftline{\bf APPENDIX D}
\leftline{\bf DYC CAMAC AND RESET PAL PROGRAM LISTING: FBCAM.ABL}
\begin{verbatim}
Module  CamacClear

Title   'Responds to Camac Z and F9 by sending a Reset pulse

         Sten Hansen       Fermilab Physics Dept.        5-3-91'

        FBCAM                device          'F153';
"       Location                               U9
"Inputs..
        F16,F8,F4,F2,F1         pin      1,2,3,4,5;
        S2,Z,N                  pin      6,7,8;

"Outputs
        Q,X,SysRst              pin      9,11,19;
        NimRst,BdRst            pin      18,17;

"Group Camac F lines into a set
 FCode = [!F16,!F8,!F4,!F2,!F1];

C,z = .C.,.Z.;

Equations

SysRst  = !N & (FCode == 9) & !S2 # !Z & !S2 # BdRst # NimRst;

!X      = !N & (FCode == 9);
 X.OE   = !N & (FCode == 9);

!Q      = !N & (FCode == 9);
 Q.OE   = !N & (FCode == 9);

Test_Vectors

([N,F16,F8,F4,F2,F1,S2,Z,NimRst,BdRst]->[Q, X, SysRst])
 [1, 1 , 1, 1, 1, 1, 1,1,  0   ,  0  ]->[z, z,    0  ];
 [1, 1 , 0, 1, 1, 0, 1,1,  0   ,  0  ]->[z, z,    0  ];
 [0, 1 , 0, 1, 1, 0, 0,1,  0   ,  0  ]->[0, 0,    1  ];
 [0, 1 , 0, 1, 1, 0, 1,1,  0   ,  0  ]->[0, 0,    0  ];
 [1, 1 , 1, 1, 1, 1, 1,0,  0   ,  0  ]->[z, z,    0  ];
 [1, 1 , 1, 1, 1, 1, 0,0,  0   ,  0  ]->[z, z,    1  ];
 [1, 1 , 1, 1, 1, 1, 1,0,  0   ,  0  ]->[z, z,    0  ];
 [1, 1 , 1, 1, 1, 1, 1,1,  1   ,  0  ]->[z, z,    1  ];
 [1, 1 , 0, 1, 1, 0, 1,1,  0   ,  1  ]->[z, z,    1  ];

end CamacClear
^Z
\end{verbatim}
\end{document}